\documentclass[twocolumn]{aastex63}

\usepackage{xcolor}
\usepackage{amsmath}
\usepackage{affils}
\usepackage{booktabs}
\newcommand{\squig}{SQuIGG$\vec{L}$E}

\newcommand{\dn}{D{\ensuremath{_{\rm{n}4000}}}}

\def\gtrsim{\mathrel{\hbox{\rlap{\hbox{\lower4pt\hbox{$\sim$}}}\hbox{\raise2pt\hbox{$>$}}}}}

\newcommand{\hbeta}{H\ensuremath{\beta}}

\newcommand{\loiii}{\ensuremath{L_{\mathrm{[O {\tiny III}]}}}}

\newcommand{\msun}{\ensuremath{M_{\odot}}}
\newcommand{\mstar}{\ensuremath{M_*}}

\newcommand{\nii}{[\ion{N}{2}]}

\newcommand{\oii}{[\ion{O}{2}]}
\newcommand{\oiii}{[\ion{O}{3}]}

\def\lesssim{\mathrel{\hbox{\rlap{\hbox{\lower4pt\hbox{$\sim$}}}\hbox{$<$}}}}
\def\gtrsim{\mathrel{\hbox{\rlap{\hbox{\lower4pt\hbox{$\sim$}}}\hbox{$>$}}}}

\begin{document}

\shortauthors{Greene et al.}

\title{The Role of Active Galactic Nuclei in the Quenching of Massive Galaxies in the SQuiGG$\vec{L}$E Survey}

\author{Jenny E. Greene}
\affiliation{Department of Astrophysical Sciences, Princeton University, Princeton, NJ 08544, USA}
\author{David Setton}
\affiliation{Department of Physics and Astronomy and PITT PACC, University of Pittsburgh, Pittsburgh, PA, 15260, USA}
\author{Rachel Bezanson}
\affiliation{Department of Physics and Astronomy and PITT PACC, University of Pittsburgh, Pittsburgh, PA, 15260, USA}
\author{Katherine A. Suess}
\affiliation{Astronomy Department, University of California, Berkeley, CA 94720, USA}
\author{Mariska Kriek}
\affiliation{Astronomy Department, University of California, Berkeley, CA 94720, USA}
\author{Justin S. Spilker}
\affiliation{Department of Astronomy, University of Texas at Austin, 2515 Speedway, Stop C1400, Austin, TX 78712, USA}
\author{Andy D. Goulding}
\affiliation{Department of Astrophysical Sciences, Princeton University, Princeton, NJ 08544, USA}
\author{Robert Feldmann}
\affiliation{Institute for Computational Science, University of Zurich, CH-8057 Zurich, Switzerland}

  \date{\today}

\begin{abstract}
We study the incidence of nuclear activity in a large sample of massive post-starburst galaxies at $z \sim 0.7$ selected from the Sloan Digital Sky Survey, and identify active galactic nuclei based on radio continuum and optical emission lines. Over our mass range of $10^{10.6} - 10^{11.5}$~\msun, the incidence of radio activity is weakly dependent on stellar mass and independent of stellar age, while radio luminosity depends strongly on stellar mass. Optical nuclear activity incidence depends most strongly on the \dn\ line index, a proxy for stellar age, with an active fraction that is $\sim$ ten times higher in the youngest versus oldest post-starburst galaxies. Since a similar trend is seen between age and molecular gas fractions, we argue that, like in local galaxies, the age trend reflects a peak in available fueling rather than feedback from the central black hole on the surrounding galaxy.
\end{abstract}

\section{Introduction}
\label{sec:intro}

Ever since the realization that supermassive black holes (BHs) are ubiquitous at the centers of massive galaxies, we have invoked energy input from active galactic nuclei (AGN) as a mechanism for heating or removing the interstellar medium of galaxies and suppressing star formation \citep[e.g.,][]{silk:98,springel:05}. While there is strong evidence that radio emission from central galaxies keeps the intracluster medium from cooling in the cluster environment \citep[e.g.,][]{fabian:12}, we do not know whether AGN play an active role in shutting off star formation in massive elliptical galaxies \citep[e.g.,][]{heckmanbest:14}. It is hard to define an experiment that can directly implicate the AGN, given that star formation necessarily lasts at least an order of magnitude longer in time than the typical mass doubling time of a BH (the Salpeter time of $3 \times 10^7$ yr). 

Large AGN surveys (predominantly identified using X-rays) have shown clearly that accretion depends on both the mass of the galaxy (and thus presumably the black hole) and the star formation rate \citep[e.g.,][]{silverman:09,aird:12,chen:13,yang:17,yang:18,aird:18,aird:19}. In cases where detailed spectroscopic information is available, there is some evidence that AGN tend to occur a few hundred million years after a burst of star formation \citep[e.g.,][]{kauffmann:03,davies:07,wild:10}, although optical spectroscopy may be biased against finding accretion activity in the most luminous starbursts \citep[e.g.,][]{trump:15,jones:16}.

In this work, we identify galaxies for which a burst of star formation has ended recently, so-called post-starburst (PSB) galaxies. We are interested in massive galaxies whose {\it primary} epoch of star formation has recently ended. Since massive galaxies ended their star formation early \citep[e.g.,][]{thomas:05}, observations at significant lookback times are required to find massive PSBs. Such galaxies are identifiable, although quite rare, at the modest redshift of $0.6<z<1$ \citep[e.g.,][]{tremonti:07,pattarakijwanich:14,wild:16}. In this work, we ask whether a sample of massive and recently quenched PSBs show any sign of enhanced nuclear activity compared with quiescent galaxies of similar mass. 
We adopt a standard concordance cosmology with $H_0 = 70$~km~s$^{-1}$~Mpc$^{-1}$, $\Omega_m = 0.3$, $\Omega_{\Lambda}=0.7$.

\begin{figure} 
\hspace{-1cm}
\includegraphics[width=10cm]{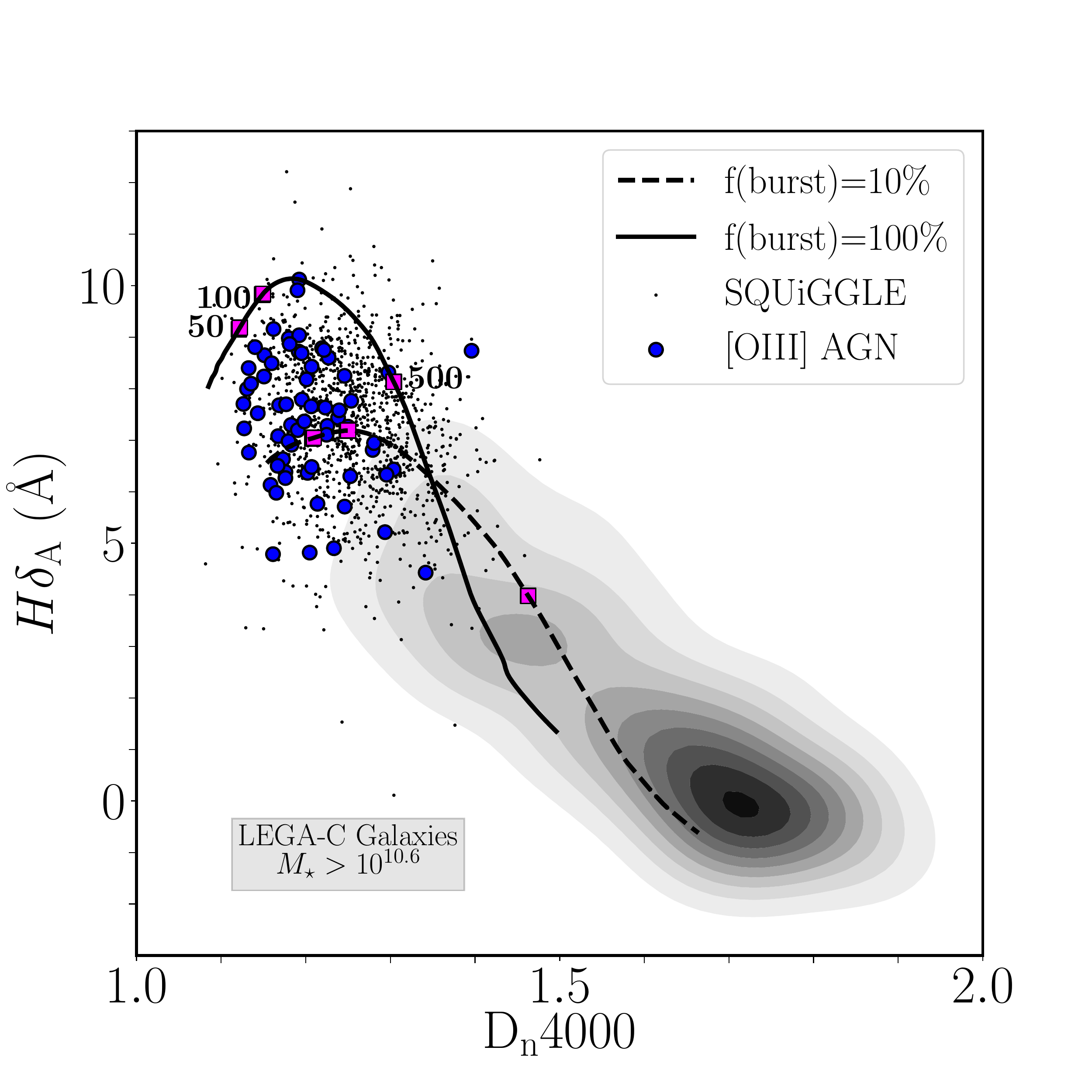}
\vspace{-2mm}
\caption{ \squig\ galaxies (dots) are selected based on their Balmer break strength, but the result is a sample with low \dn\ and high H$\delta_{\rm A}$. The \oiii-selected AGN are highlighted in blue dots. In comparison, the LEGA-C mass-selected sample shows how a mass-selected sample of galaxies move in this two-dimensional plane. The two star-formation history tracks quantify the young ages of the sample. These two-burst star-formation histories have either 10\% or 100\% of the mass formed in a recent burst with ages 50, 100, 500 Myr as shown with magenta squares.}
\label{fig:hddn}
\vspace{1pt}
\end{figure}

\section{Post-Starburst Sample}
\label{sec:sample}

\begin{figure*} 
\begin{minipage}{0.5\linewidth}
\includegraphics[width=9.0cm]{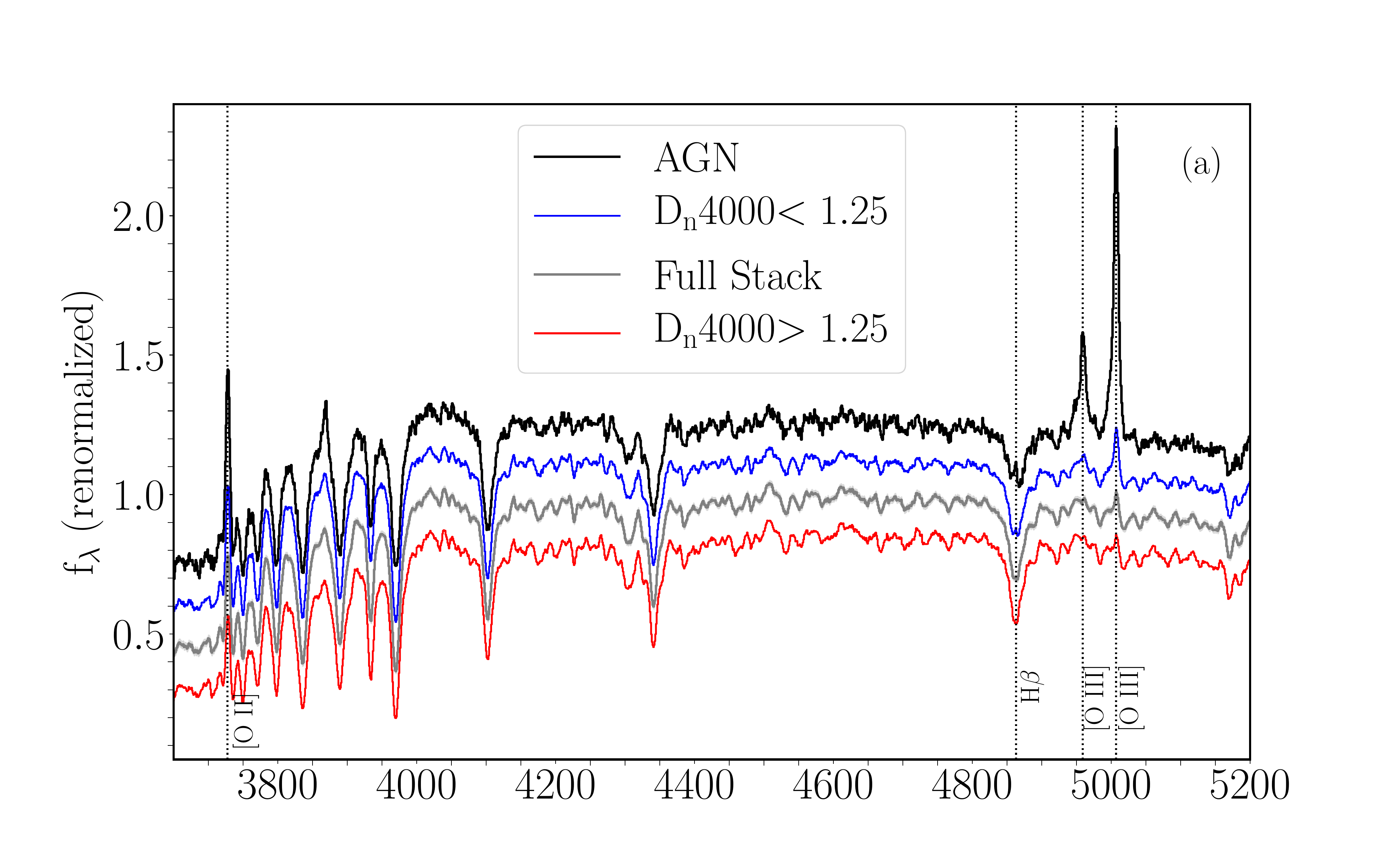} 
\end{minipage}
\hspace{-20mm}
\begin{minipage}{0.5\linewidth}
\includegraphics[width=5.6cm]{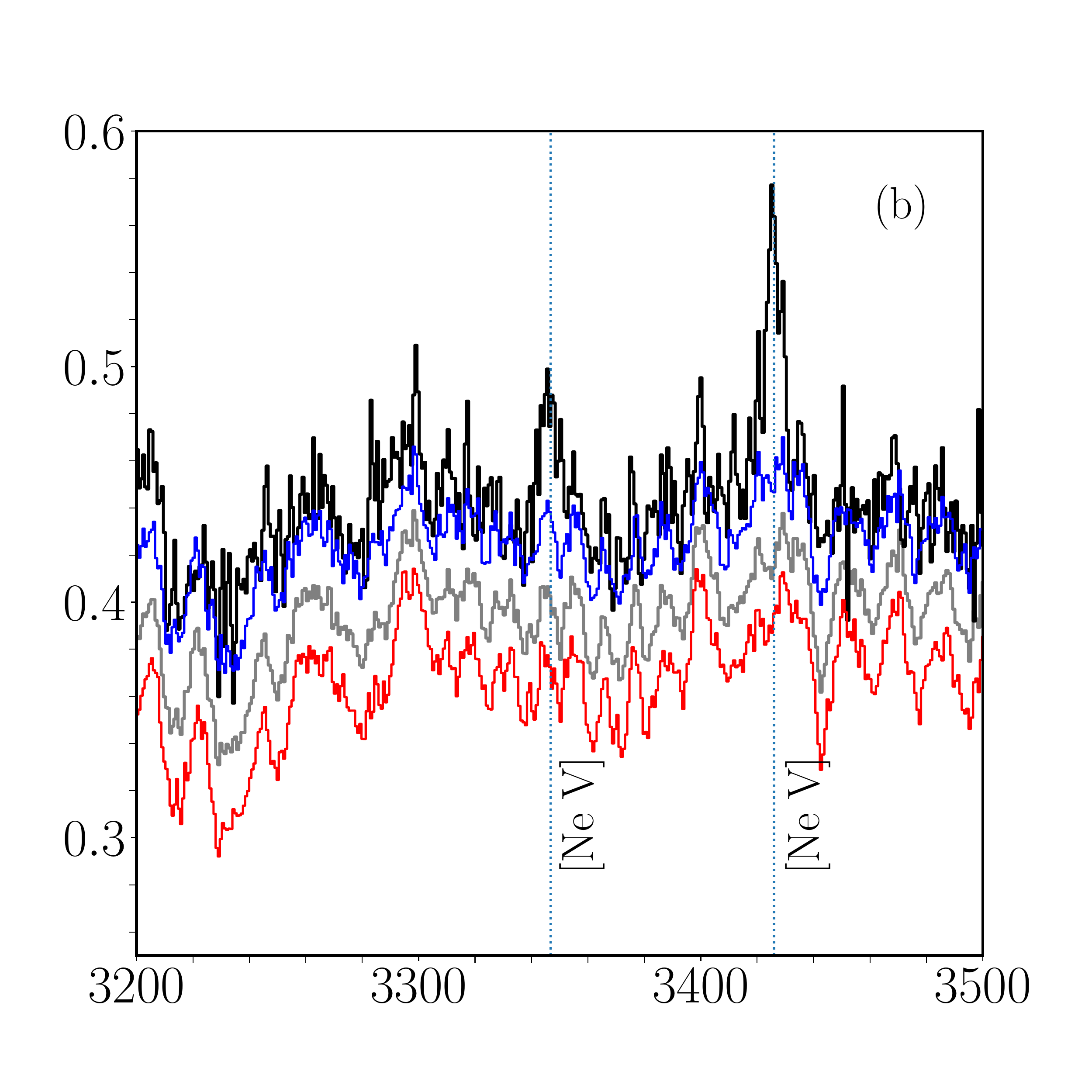} 
\end{minipage}
\begin{minipage}{0.5\linewidth}
\centering
\includegraphics[width=9.0cm]{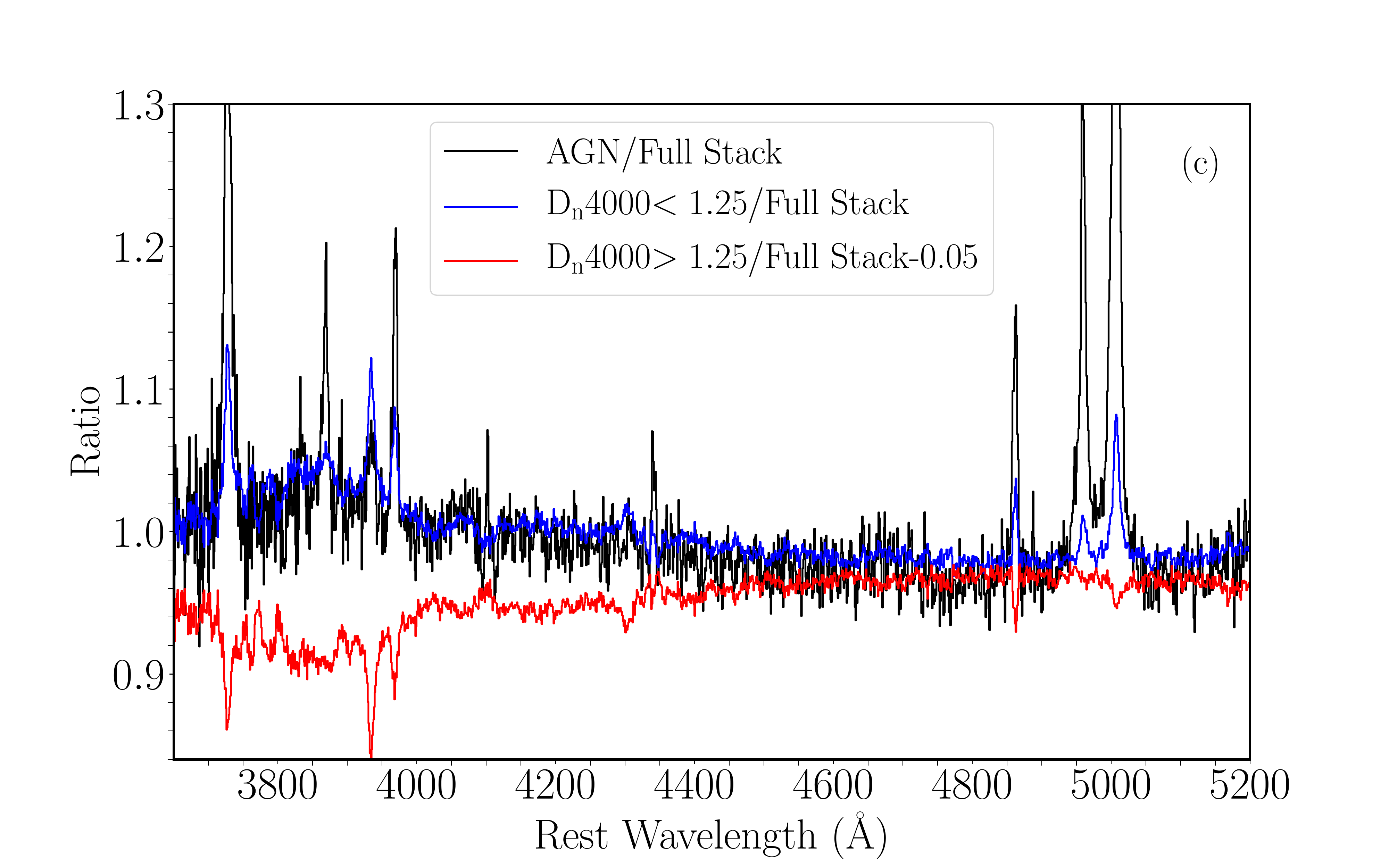} 
\end{minipage}
\hspace{-2.5mm}
\begin{minipage}{0.5\linewidth}
\includegraphics[width=5.6cm]{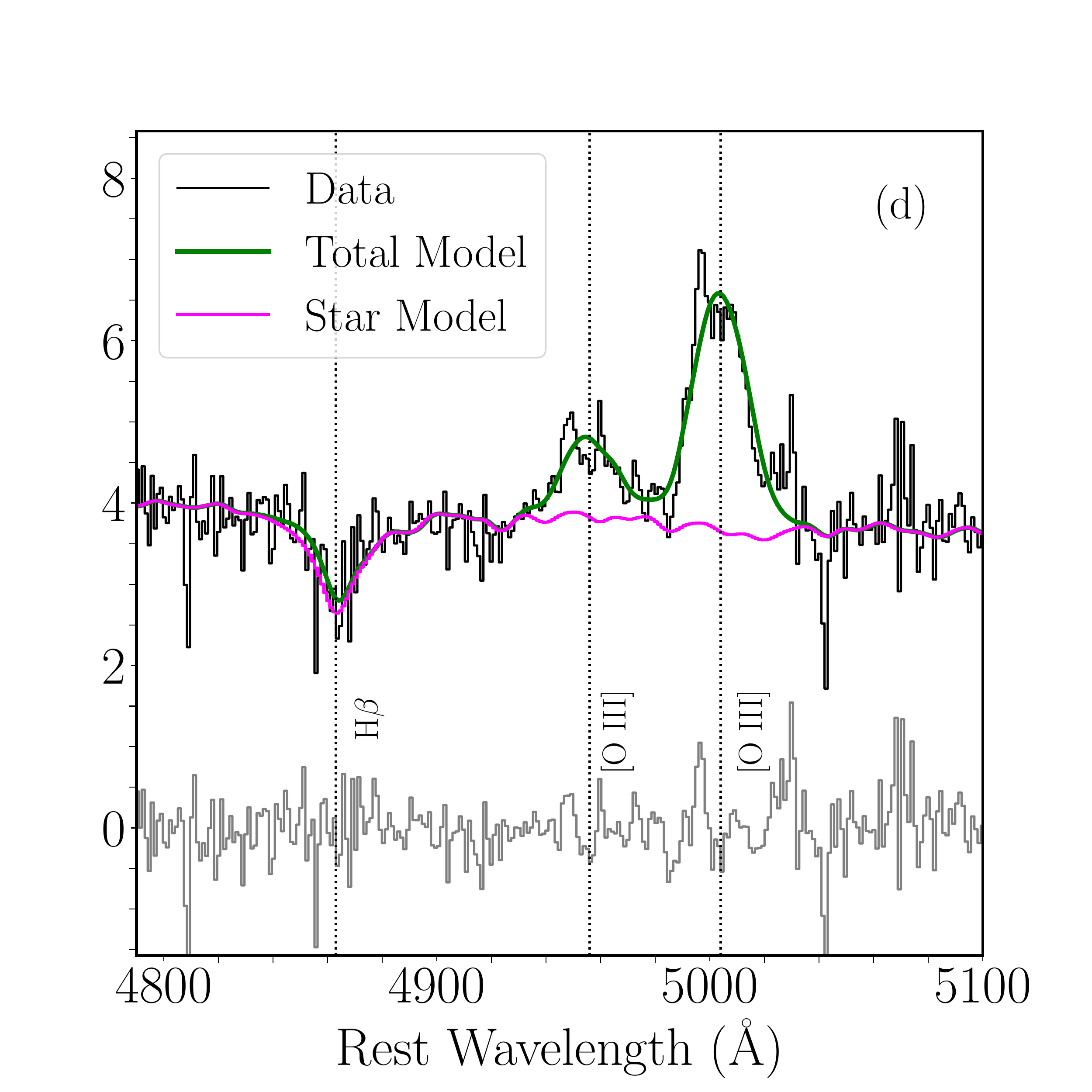}
\end{minipage}
\caption{(a) Composite spectrum for the 64 galaxies identified as AGN based on their \oiii/H$\beta$ ratio (black), for the full sample of 1207 \squig\ galaxies in CMASS (blue, 1$\sigma$ spread in grey), and for the non-AGN with \dn$<1.25$ (orange). Small offsets applied for clarity. (b) A zoom-in of the [\ion{Ne}{5}]$\lambda \, \lambda 3347, 3426$ lines, that are only visible in the AGN stack. (c) With full stack in the denominator, ratio of AGN (black), low \dn\ (blue), and high \dn\ (red) galaxies, showing that the stellar continua are similar within a few percent and highlighting differences in emission lines. (d) The \hbeta+\oiii\ region for a single object, along with the pPXF fit (magenta) and emission lines (green).}
\label{fig:stack}
\vspace{5pt}
\end{figure*}

In order to understand the detailed processes driving galaxy quenching, we have launched the Studying of QUenching in Intermediate-z Galaxies: Gas, angu$\vec{L}$ar momentum, and Evolution (\squig) Survey. In Suess et al.\ in preparation, we present a large new sample of 1,318 post-starburst galaxies that are selected from the Sloan Digital Sky Survey Data Release 14 \citep[SDSS DR14;][]{dr14}. The sample ranges in redshift from $z = 0.5 - 0.94$, (median $z=0.68$). The galaxies are relatively bright and massive, with $i=17.9 - 20.5$ mag (median $i=19.5$ mag), with nearly all having $10.5 < \log M_*/\msun < 11.5$ (median $M_* = 10^{11}$~\msun). 

The sample was selected from the SDSS DR14 spectroscopic catalog following \citet[][]{kriek:10}. In brief, we calculate the flux in synthetic rest-frame $U$, $B$, and $V-$band filters for all SDSS spectra with $z>0.5$. We require that the spectra have S/N $>= 6$ in both the $U$ and $B$ filters and then use the color cuts $U-B > 0.975$ and $-0.25 < B-V < 0.45$ to isolate galaxies with the unique A-star--like spectral shape indicative of a recently-quenched stellar population.

To demonstrate the success of our selection, we look at two key stellar absorption indices (Fig.\ \ref{fig:hddn}). The equivalent width of H$\delta_{\rm A}$ combined with the \dn\ index (the ratio in flux at 3850-3950\AA\ with that at 4000-4100\AA) are a powerful non-parametric way to track stellar age \citep[e.g.,][]{kauffmann:03}. We show two tracks in Figure \ref{fig:hddn} based on two-component stellar population models with an old population plus a young burst with varying burst fraction. \squig\ galaxies had bursts of star formation that ended 100-500 Myr in the past. The actual quenching time is degenerate with the burst fraction, but the light-weighted ages tend to be $\sim 500$~Myr for a \dn$=1.25$ (D. Setton et al.\ in preparation). 

Within SDSS, objects are targeted spectroscopically for many reasons, including radio emission. To avoid any pre-selection based on active galaxy attributes, we restrict attention to the 1207 galaxies that fall within the ``CMASS'' SDSS sample. This sample is targeted based solely on $griz$ color and magnitude \citep{dawson:13}. The inverse-variance--weighted mean spectrum of the full sample, normalized at 4020\AA, is shown in Figure \ref{fig:stack}. 

\squig\ aims to study the gas fractions, angular momenta, stellar populations, and merger fractions of this PSB sample. With ALMA, we have searched for CO 2-1 in 13 galaxies \citep{suess:17}, and found that roughly half contain surprisingly large gas reservoirs, with gas fractions of $\sim 10-20\%$ (R.\ Bezanson et al.\ in preparation). 

\begin{figure*} 
\hspace{-1cm}
\includegraphics[width=10cm]{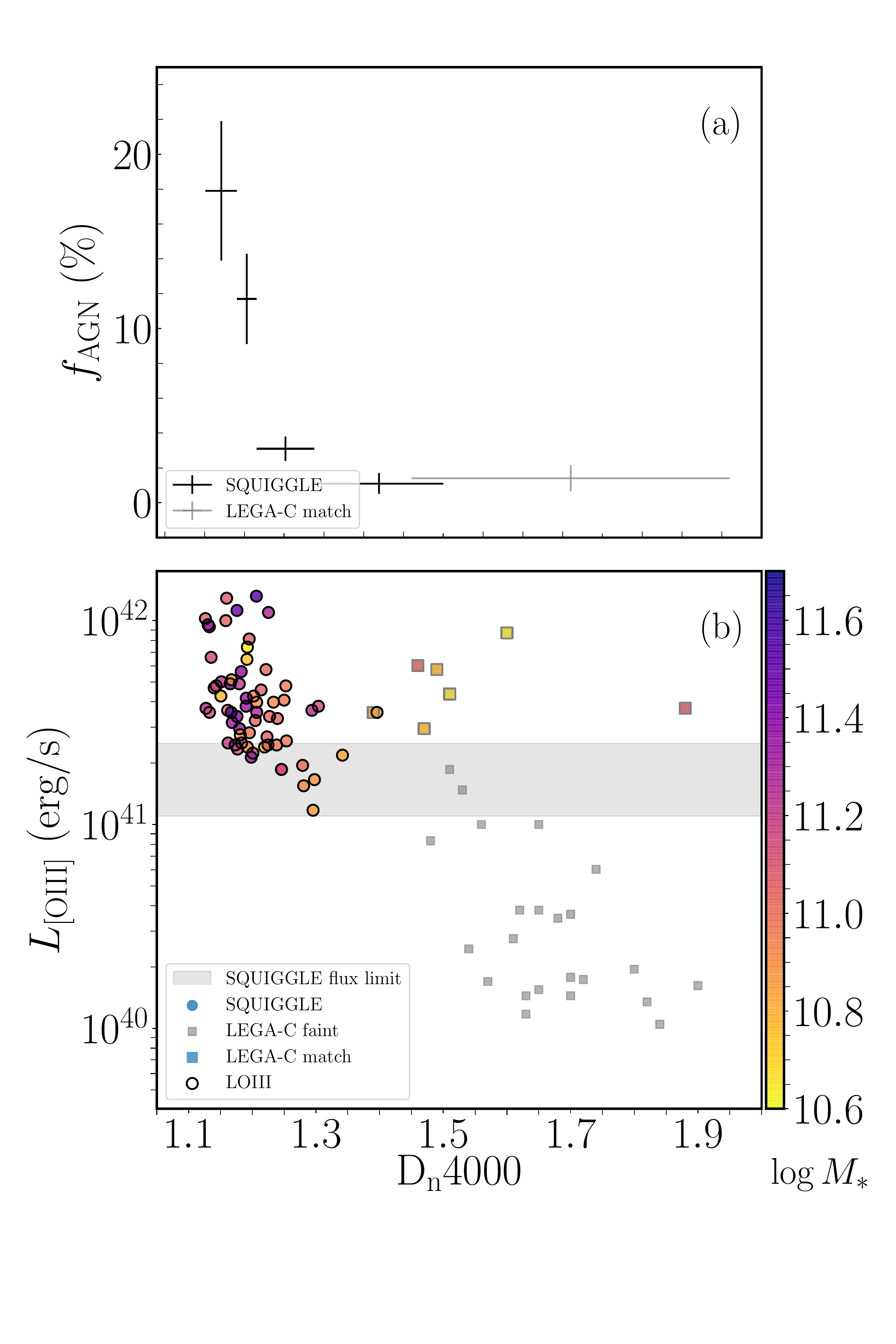}
\hspace{+0cm}
\includegraphics[width=10cm]{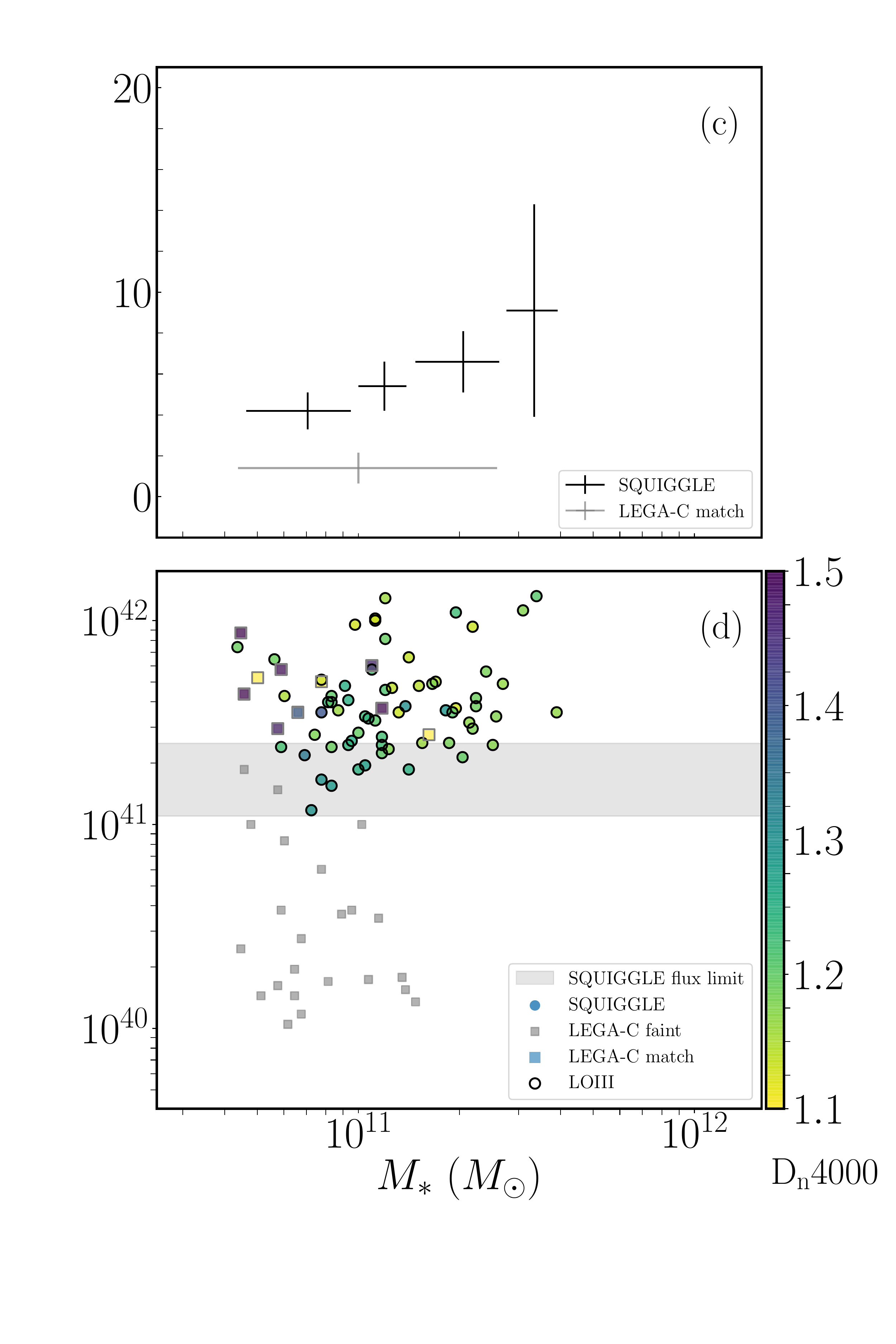}
\vspace{-1.5cm}
\caption{The relationship between galaxy properties and the \oiii\ luminosity, as a proxy for the AGN luminosity. In addition to the \squig\ sample (circles) we include the LEGA-C sample for comparison (squares). The LEGA-C galaxies are selected in the rest-frame optical and so span the full range of morphologies and star-formation histories of galaxies in a matching mass range (mostly being quiescent). LEGA-C AGN below the \squig\ flux limit (which translates into the grey band in terms of luminosity) are shown in grey squares. On the left, we examine the AGN fraction (a) and the \oiii\ luminosity (b) as a function of \dn, a proxy for stellar age, with younger galaxies further left. There is a dramatic rise of $\sim$ ten times in the AGN fraction towards the youngest ages (a), and a corresponding increase in \oiii\ luminosity at young ages among the \squig\ galaxies. On the right, in contrast, we see at most a weak trend between AGN fraction (c) or \oiii\ luminosity (d) and galaxy stellar mass. }
\label{fig:oiii}
\vspace{1pt}
\end{figure*}

\subsection{LEGA-C Comparisons}

We rely on the Large Early Galaxy Astrophysics Census \citep[LEGA-C,][]{wel:16} sample to provide a more complete view of the galaxy population at this epoch. The LEGA-C survey is an ESO Large spectroscopic program, which consists of deep (20-hour) integrations of $\sim3500$ $K$-magnitude selected galaxies at $0.6\lesssim z \lesssim1$, regardless of morphology or color, in the COSMOS field. Thus, we can mitigate possible bias from the CMASS color selection by comparison with LEGA-C. Although we will not take full advantage of their exquisite spectra, the color-independent selection provides a mass-selected sample of galaxies to compare with \squig. While covering a larger area on the sky, the CMASS sample is very incomplete in our mass and redshift range of interest \citep{leauthaud:16}.  Furthermore, the deep 3 GHz VLA imaging in COSMOS \citep{smolcic:17} and resulting radio source catalog for LEGA-C \citep{barisic:17}, allows us to identify AGN in both LEGA-C and \squig\ using similar radio and optical diagnostics.

We use the publicly available LEGA-C Data Release 2 \citep{straatman:18}, and select galaxies in a matching mass range of $10.6 < \log M_*/\msun < 11.5$. The stellar masses of both samples are similarly determined using the Fitting and Assessment of Synthetic Templates \citep[FAST;][]{kriek:09} code, assuming the same initial mass function \citep{chabrier:03} and stellar population models \citep{bc:03} (and in the case of LEGA-C the UltraVISTA photometry of \citealt{muzzin:13ultravista}). We select our LEGA-C comparison sample based only on mass, regardless of stellar age or galaxy morphology. The redshift ranges are similar enough that we do not specifically match them. 

\section{Spectral Fitting}

We select AGN using the ratio between \oiii$\lambda 5007$ and H$\beta$, requiring reliable measurements of the emission line strengths of these two key lines. Due to the strong stellar absorption in H$\beta$, we jointly model the stellar continuum and emission lines (\oii$\,\lambda\lambda 3726, 3729$, H$\beta$, and \oiii$\,\lambda\lambda 4959,5007$) using the publicly available penalized pixel-fitting code pPXF \citep[][]{cappellari:04}. Our stellar templates are the single stellar population models from \citet{vazdekis:2010} built from the MILES stellar library \citep{miles}, and we include an order 12 additive polynomial to account for flux calibration uncertainties. We model the line-of-sight stellar velocity distribution as a Gaussian and fix the velocities and linewidths of all emission lines to the same value. 

We generate mock spectra to measure uncertainties on our flux measurements. We start with the best-fit pPXF model fit to the stars and gas, and for each of 100 artificial runs we draw the noise spectrum randomly from the variance array, and refit. The standard deviation in flux values from these runs is our adopted uncertainty, which is 20\% higher than the statistical errors returned by pPXF. We select all galaxies with \oiii\ line equivalent width EW$>5$, which corresponds to a minimum S/N$=4$ in the line, although all but ten of the galaxies have S/N$>10$. H$\beta$ is most challenging to detect, given the strong absorption. In the case of non-detections of H$\beta$ emission, we assign the $3 \sigma$ upper limit as the flux for the purpose of calculating line ratios.

\section{Identifying AGN}
\label{sec:idagn}

We select two complementary AGN samples. First, we look for high-ionization optical emission line ratios indicative of a hard ionizing spectrum. Then, we use the Faint Images of the Radio Sky at Twenty-cm \citep[FIRST;][]{becker:95} survey to find AGN through radio emission. We adopt the FIRST survey because it overlaps so well with the SDSS footprint, and it remains the most uniform survey available over this area with a beam size roughly matched to that of galaxies.

\subsection{Optical Emission-line Selection}

We search for high-ionization emission lines in the SDSS PSB spectra. Classic narrow-line region selection relies on two-dimensional emission-line ratio plots known as BPT diagrams \citep{baldwin:81}. Unfortunately, at the redshifts of our sample, the SDSS spectra do not include H$\alpha$ or \nii~$\lambda \lambda 6548, 6584$, which play a central role in standard BPT diagrams. Instead, we are forced to rely on \oiii/H$\beta$ alone. 

At first glance, this single line ratio does not appear sufficient to classify a galaxy as an AGN. In standard BPT diagrams, there are many galaxies without nuclear activity but with high values of \oiii/\hbeta. However, in the absence of AGN activity, high \oiii/\hbeta\ is predominantly seen in low-mass, low-metallicity galaxies \citep[e.g.,][]{moustakas:06}. At higher masses, star-formation--powered emission lines have lower \oiii/\hbeta\ ratios. This realization was formalized by \citet{juneau:11} in what they call the Mass-Excitation (MeX) diagram, which plots \oiii/\hbeta\ against stellar mass, and shows that local galaxies with $M_* > 10^{10.8}$~\msun\ and \oiii/\hbeta$>3$ are all AGN. While strong line ratios are known to shift within the BPT diagram at higher redshift \citep[e.g.,][]{shapley:05}, in our redshift range of interest $z \sim 0.7$, galaxies are observed to have ISM emission lines similar to local galaxies \citep[e.g.,][]{kewley:13}. Furthermore, even at $z \sim 2$, higher-mass galaxies like those studied here do not appear significantly offset \citep[e.g.,][]{shapley:15}.   

We apply the simple joint criteria that (a) the \oiii\ line EW$>5$\AA\ and (b) \oiii/\hbeta$>3$ \citep[e.g.,][]{ho:08}. This ensures adequate signal-to-noise in the line and corresponds to a flux limit of $10^{-16}$erg~s$^{-1}$~cm$^{-2}$. All but four of the galaxies with \oiii\ above our EW cut qualify as AGN based on line ratios. These two cuts identify a total of 64 AGN, out of 1207 galaxies, for an AGN fraction of $5 \pm 0.7 \%$ (see Table 1). As additional confirmation that we are selecting bona-fide AGN, we note that the \oiii/\hbeta\ distribution is peaked at a ratio of 10, and that our qualitative results would not change if we selected the 50 objects with \oiii/\hbeta$>5$.

In Figure \ref{fig:oiii}, we plot the \oiii\ luminosity distribution of the detected AGN. Our objects span only one order of magnitude in \oiii\ luminosity. Sources with fluxes $< 10^{-16}$~erg s$^{-1}$~cm$^{-2}$ or $L_{\rm [OIII]} < 10^{41}$~erg~s$^{-1}$ at $z=0.7$ are not included in our sample. We are likely less complete in AGN at lower stellar masses, given the correlation between \loiii\ and $M_*$. We quantify this correlation and investigate the impact of this incompleteness in \S \ref{sec:oiiidemo}. 
Furthermore, the most luminous targets may be too blue to either be included in the CMASS sample or the PSB color selection, due to scattered light from the obscured quasar \citep[e.g.,][]{zakamska:06,liu:09} but such objects are very rare \citep{reyes:08}. For physical context, black holes of $M_{\rm BH} \approx 10^8$~\msun\ \citep[for typical stellar masses of $\sim 10^{11}$~\msun;][]{haeringrix:04} have a range in Eddington ratio of 1-10\%, assuming a bolometric correction of 10$^3$ \citep{liu:09} for \oiii. 

Through stacking, we compare the average spectra of the \oiii-selected AGN with the full inactive sample, as well as stacks of the younger and older galaxies divided at the median \dn$=1.25$. We can see that the AGN have a bluer stellar continuum than the full stack, matching that of the younger inactive stack (\S \ref{sec:oiiidemo}), but the line emission is much stronger in the AGN stack than in the young inactive stack. We also detect [\ion{Ne}{5}]$\lambda 3426$~\AA, a line that is very hard to excite without the hard ionizing spectrum of an AGN \citep[e.g.,][]{goulding:09}. In Figure \ref{fig:stack}b, we show that the [\ion{Ne}{5}] line is clearly detected in the stacked AGN spectrum, providing some confidence that a significant fraction of our optically selected AGN are powered by accreting black holes.

\subsubsection{LEGA-C Comparison Sample}

We select as LEGA-C AGN those galaxies with \oiii/\hbeta$>3$ and an \oiii\ flux above the \squig\ \oiii\ flux limit of $10^{-16}$~erg~s$^{-1}$. The resulting five galaxies are shown as colored squares in Figure \ref{fig:oiii} corresponding to an AGN fraction of $1.4 \pm 0.3$ \%. We also relax the \oiii\ luminosity cut and show all galaxies with AGN-like line ratios in LEGA-C (grey squares). Unlike the \squig\ sample, LEGA-C does not contain many massive PSBs, and it has very few AGN above our flux limit. Both are almost certainly due to the small volume.

\subsection{Radio Selection}

To find the radio-detected AGN within the \squig\ sample, we search the FIRST archive for cross-matches within 5\arcsec\ of the position of each \squig\ galaxy. In practice, all but two are matched within 3\arcsec\ of the SDSS position. One of those is a extended jet source, leaving one possible mis-matched source. We do not inspect the images for possible radio detections at larger radius \citep[as in, for instance,][]{best:05}. For ease of comparison with other radio AGN samples, we calculate our radio detection limit at 3 GHz rather than 1.4 GHz, assuming a standard radio spectral slope of $\alpha = -0.7$. FIRST has a uniform flux limit of 1 mJy, which corresponds to a luminosity limit range $L_{\rm 3 GHz} = 0.6-2 \times 10^{24}$~W~Hz$^{-1}$, and $L_{\rm 3 GHz} = 10^{24}$~W~Hz$^{-1}$ for a median distance of $z=0.7$.

\begin{figure*} 
\hspace{-1cm}
\includegraphics[width=10cm]{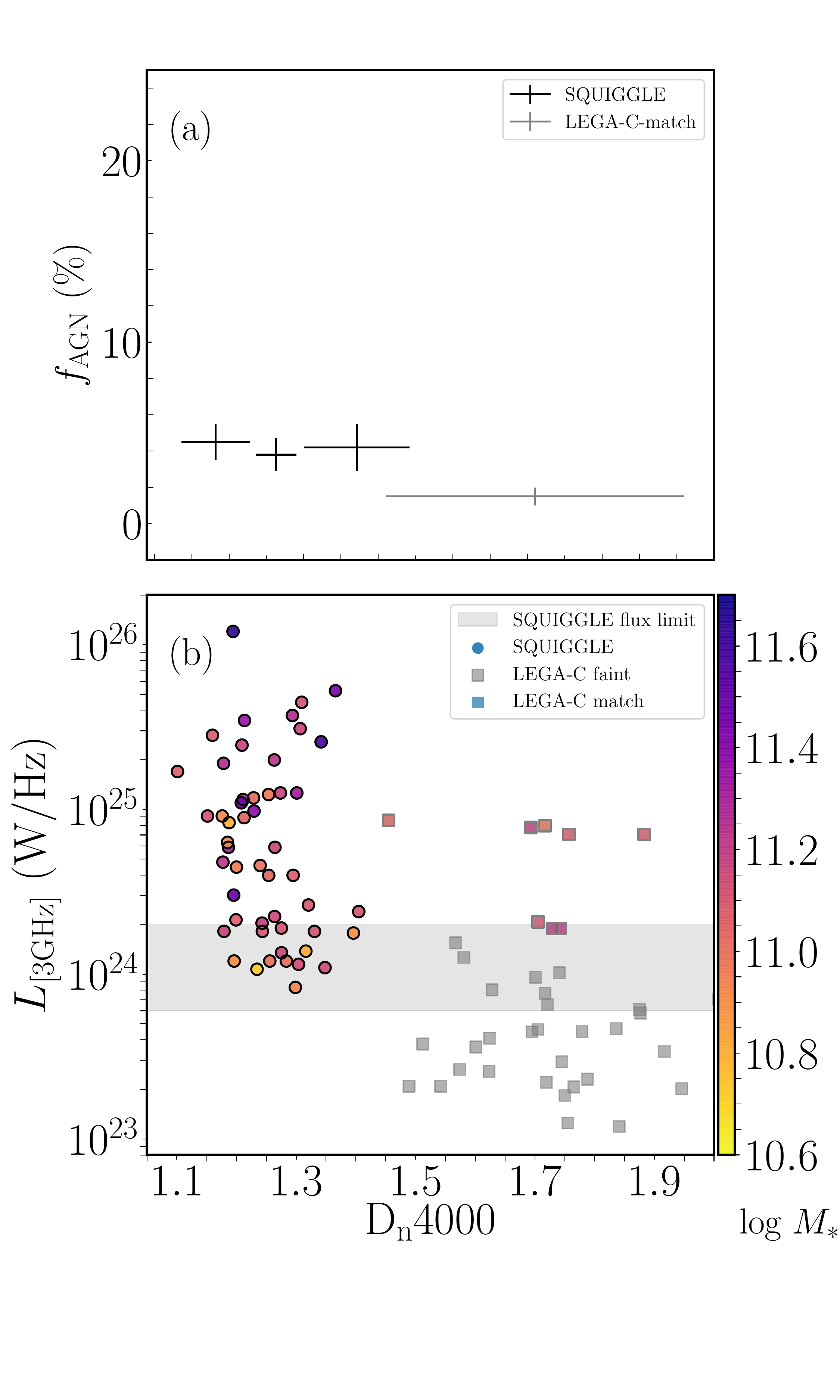}
\includegraphics[width=10cm]{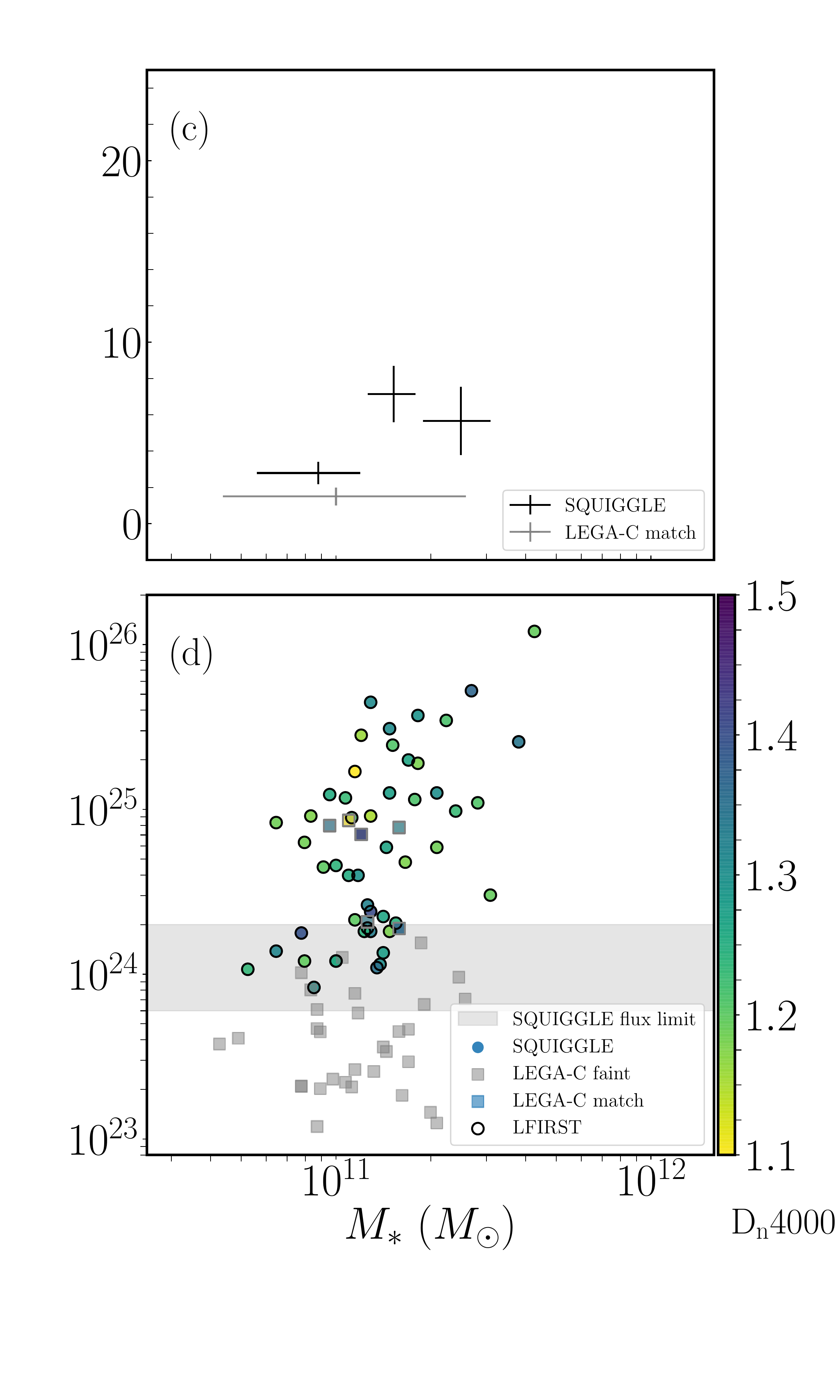}
\vspace{-1.5cm}
\caption{The relationship between galaxy properties and luminosity for radio AGN in \squig\ (circles, black crosses) and LEGA-C (squares, grey crosses), including LEGA-C radio AGN below our flux limit of $f_{\rm 1.4 GHz} = 1$~mJy (grey squares). On the left, we examine the AGN fraction (a) and the 3 GHz luminosity (b) as a function of \dn, a proxy for stellar age. Younger galaxies have lower \dn. We see no relationship between radio luminosity and stellar age, in contrast to \oiii\ luminosity. On the right, we see a strong trend between radio luminosity and galaxy mass (d), but only a weak trend between AGN fraction and galaxy stellar mass (c).
}
\label{fig:radio}
\vspace{1pt}
\end{figure*}

This selection yields 52 radio AGN, or a radio AGN fraction of $4 \pm 0.6$\%. The distribution of radio luminosities are plotted in Figure \ref{fig:radio}. The star formation rate needed to explain the observed radio emission is many hundreds of solar masses per year \citep[e.g.,][]{bell:03}. Given that we observe upper limits on the star formation rates of tens of solar masses per year from \oii\ fluxes \citep{suess:17}, we view as highly unlikely that the radio emission we observe arises from star formation.

In contrast to the \oiii-selected sample, we observe a very wide range of radio luminosities, spanning nearly three decades (Figure \ref{fig:radio}). Radio luminosity does not have a simple or well-understood relationship with the accretion rate \citep[see references in][]{heckmanbest:14}, although radio emission contributes a larger fraction of the bolometric luminosity at lower AGN accretion rates \citep[e.g.,][and references therein]{ho:08}. We detect \oiii\ emission from only five of our radio galaxies. The majority appear to be ``low-excitation'' \citep[][]{hinelongair1979} sources that, likely due to very low Eddington ratios, do not show radiative signatures of standard optically thick but geometrically thin accretion disks \citep[e.g.,][]{shakurasunyaev1973,bestheckman2012}. Conversely, among the \oiii\ sources, we detect few radio AGN due to our flux limit; \oiii-selected AGN in SDSS with $L_{\rm [OIII]} \approx 10^{41}$erg~s$^{-1}$ have radio luminosities of $L_r \approx 10^{22.5}-10^{23}~$W~Hz$^{-1}$ as probed by stacks \citep[e.g.,][]{deVries:07}.

\subsubsection{LEGA-C Comparison}

\citet{barisic:17} built a catalog of radio AGN in LEGA-C; we simply take the subset of these that obey our flux limit of $f_{\rm 1.4 GHz} = 1$~mJy and fall within our stellar mass range. We find a radio AGN fraction of $1.5 \pm 0.4$\%. In the case of radio AGN, the fraction in \squig\ is consistent within $2 \sigma$ of the radio-loud fraction in a mass-matched LEGA-C sample.

\section{AGN Demographics}

Having identified a sample of recently-quenched galaxies that host AGN, we now examine trends between the AGN and host galaxy properties. As shown in Fig.\ \ref{fig:hddn}, \dn\ tracks stellar age; the median \dn$\,=1.25$ corresponds to a light-weighted age of $\sim 500$~Myr. In the following, binned quantities are constructed to have equal numbers of AGN per bin, and error bars are Poisson.

\subsection{\oiii-selected AGN}
\label{sec:oiiidemo}

We see a dramatic rise in AGN fraction towards lower values of \dn\ (Fig.\ \ref{fig:oiii}). AGN are ten times more likely to occur in the youngest as in the oldest \squig\ galaxies, rising from $1 \pm 0.6$\% to $17 \pm 4$\% from \dn$=1.3$ to $1.1$. Interestingly, all of the \squig\ galaxies with ALMA detections are similarly in galaxies with \dn$\lesssim 1.3$.

In contrast, we see a very weak trend between galaxy stellar mass and AGN fraction (Figure 2cd). Despite the weak trend between AGN fraction and stellar mass, we find a strong correlation between \loiii\ and $M_*$ for the detected AGN. Using a partial pearson correlation analysis, we find $P_{\loiii, D_{\rm n}4000} = 0.5$ and $P_{\loiii, M_*} = 0.9$. The trend between \loiii\ and $M_*$ is shallow, but highly significant.

The lower \mstar\ objects likely suffer more \oiii\ flux incompleteness. To investigate the impact of preferentially missing AGN in lower-\mstar\ galaxies, we measure the AGN fraction as a function of \dn\ and \mstar\ for (a) galaxies with $z<0.7$, that should be more complete in \loiii\ and (b) galaxies with $M_* > 10^{11}$~\msun\ that should be complete in \mstar. In both cases, our primary result holds: there is an order of magnitude rise in AGN fraction over the observed range in \dn\ and no significant trend in AGN fraction with \mstar.  

\begin{deluxetable}{ccccc}
\tablecolumns{5}
\tablewidth{0pt}
\tablecaption{AGN Fractions \label{table:agnfrac}}
\tablehead{
\colhead{Sample} & \colhead{Subset} & \colhead{N$_{\rm Tot}$} & \colhead{N$_{\rm AGN}$} & \colhead{f$_{\rm AGN}$}}
\startdata
Optical & All & 1207 & 64 & $0.053 \pm 0.007$  \\
Optical & Young & 603 & 53 & $0.088 \pm 0.012$  \\
Optical & Old & 604 & 11 & $0.018 \pm 0.005$  \\
Optical & Low Mass & 602 & 26 & $0.043 \pm 0.008$  \\
Optical & High Mass & 605 & 38 & $0.063 \pm 0.010$  \\
LEGA-C & Optical & 803 & 10 &  $0.012 \pm 0.004$ \\
Radio & All & 1207 & 51 & $0.042 \pm 0.006$  \\
Radio & Young & 603 & 26 & $0.043 \pm 0.008$  \\
Radio & Old & 603 & 38 & $0.063 \pm 0.010$  \\
Radio & Low Mass & 602 & 14 & $0.023 \pm 0.006$  \\
Radio & High Mass & 605 & 37 & $0.061 \pm 0.010$  \\
LEGA-C & Radio & 675 & 10 &  $0.015 \pm 0.005$ \\
\enddata
\tablecomments{AGN fraction over galaxy parameters. We divide old vs. young at \dn$=1.25$ and high vs. low mass at $M_* = 10^{11}$~\msun.}
\end{deluxetable}

\subsection{Radio-Selected Sample}
\label{sec:radiodemo}

The radio sample contrasts with the \oiii-selected sample in that neither the radio luminosities nor the detection fraction shows any relation with \dn\ over our narrow range. While \citet{barisic:17} do find that radio AGN in LEGA-C are more common in older galaxies, this effect may be driven by stellar mass, since we observe a very steep relationship between stellar mass and radio luminosity within the AGN population. The partial Pearson correlation coefficients in this case are $P_{L_{3{\rm GHz}}, D_{\rm n}4000} = -0.15$ and $P_{L_{3{\rm GHz}}, M_*} = 0.9$.

This steep dependence of radio luminosity on stellar mass is well-documented in the literature for massive elliptical galaxies. \citet{heckmanbest:14} quote a slope of $L_r \propto M_*^{2.5}$. Our data also favor a super-linear slope ($L_r \propto M_*^{1.5}$) but we caution that our mass range is quite limited. It is not clear what drives this strong mass dependence, but there is clear evidence that halo mass matters. At fixed stellar mass, galaxies in more massive halos show higher radio AGN fractions \citep{mandelbaumetal2009,sabater:13}. This halo mass dependence has been ascribed to increased BH spin, a higher gas accretion rate, or jet enhancement, but the origin is unknown. We show here that stellar age is not a strong factor.

\section{Discussion and Summary}
\label{sec:discussion}

Suess et al.\ in preparation have assembled a large sample of massive and rare post-starburst galaxies at $z \sim 0.5-0.9$ using SDSS DR14 spectroscopy. We study the nuclear activity of the \squig\ sample using both optical narrow emission-line diagnostics and radio emission from FIRST. The radio AGN are likely dominated by systems that are accreting at very low Eddington fractions and are thus unobservable with optical diagnostics, while the optically selected AGN are accreting at a few percent of Eddington. 

We replicate the previously-observed trend between radio luminosity and stellar mass in radio AGN that has been reported previously \citep[e.g.,][]{best:05,barisic:17}, extending this finding to recently quenched galaxies. Radio AGN fractions are not a strong function of \mstar\ or stellar age. AGN fractions are not a strong function of \mstar\ for \oiii-selected samples either, but the optical AGN fraction is a strong function of \dn. The youngest PSBs are ten times more likely to harbor an optical AGN than the oldest PSBs. These younger galaxies, particularly those with \dn$<1.3$, are also most likely to harbor significant molecular gas fractions. What may this tell us about the relationship between black hole and galaxy growth?

Temporal connections between merging, star-formation, black hole growth, and AGN-driven outflows have long been considered in the literature \citep[e.g.,][]{sanders:88,canalizo:01,springel:05}. We note also that there is an ongoing debate about the relative roles of stellar mass and star formation rate on accretion rate distributions \citep[e.g.,][]{yang:18,aird:19}. Since we consider a very narrow range in stellar mass and redshift, we will focus only on the possible connections between star formation and accretion here. Because of the very short timescales for accretion compared with star formation, establishing an observational temporal link is challenging. Even if trends are observed, the connection could be due to either fueling or feedback. 

A number of local studies of \oiii-selected AGN have reported a boost in AGN fractions $\gtrsim{600}$ Myr after a starburst \citep[e.g.,][]{kauffmann:03,davies:07}. These studies suggest that gas availability links AGN and star formation, perhaps with winds from massive stars providing the fuel for the AGN activity \citep{wild:10}. Interestingly, \citet{aird:19} find an increase (by 2-5 times) in the AGN fraction for X-ray selected AGN in galaxies that live below the star forming main sequence as compared with galaxies on the main sequence. These results all may point to a similar fueling mechanism. We are also seeing a dramatic increase in availability of molecular gas at the lowest \dn\ in our sample. We thus postulate that the increase in AGN activity towards younger ages is a function of increased gas availability leading to fueling, rather than AGN feedback shutting down star formation in older systems. Of course, both could be true. Thus, we lastly consider whether we might link AGN activity to quenching in these systems.

Crudely speaking, theorists often conceptually divide the feedback roles of low and high accretion rate sources into a ``maintainence'' (or ``radio'') mode, and a radiatively efficient quenching mode \citep[e.g.,][]{croton:06}. The former happens in a quasi-continuous fashion to keep hot gas envelopes from cooling \citep[e.g.,][]{fabian:12}. We see clear evidence for such activity in our galaxies, and if our radio observations were deeper we would undoubtably uncover a very high radio active fraction. The latter is more elusive; despite some evidence for massive outflows in luminous AGN \citep[e.g.,][]{greene:12,liu:13}, we still do not have a clear verdict on the role of AGN in shutting down star formation in galaxies. In \squig, we will search for signs of outflow in IFU data \citep{hunt:18}. So far, we have found one intriguing object showing extended ionized and molecular gas (J.\ Spilker in preparation). We will see if there are compelling signs of outflow in this or other targets, to more cleanly disentangle the relationship between fueling and feedback in massive galaxies.

\acknowledgements We thank the referee for a very thorough report that improved this manuscript. JEG, RSB, and DN gratefully acknowledge support from NSF grant AST-1907723. RF acknowledges financial support from the Swiss National Science Foundation (grant no 157591). This work was performed in part at the Aspen Center for Physics, which is supported by National Science Foundation grant PHY-1607611.

Funding for SDSS-III has been provided by the Alfred P. Sloan Foundation, the Participating Institutions, the National Science Foundation, and the U.S. Department of Energy Office of Science. The SDSS-III web site is http://www.sdss3.org/.

SDSS-III is managed by the Astrophysical Research Consortium for the Participating Institutions of the SDSS-III Collaboration including the University of Arizona, the Brazilian Participation Group, Brookhaven National Laboratory, Carnegie Mellon University, University of Florida, the French Participation Group, the German Participation Group, Harvard University, the Instituto de Astrofisica de Canarias, the Michigan State/Notre Dame/JINA Participation Group, Johns Hopkins University, Lawrence Berkeley National Laboratory, Max Planck Institute for Astrophysics, Max Planck Institute for Extraterrestrial Physics, New Mexico State University, New York University, Ohio State University, Pennsylvania State University, University of Portsmouth, Princeton University, the Spanish Participation Group, University of Tokyo, University of Utah, Vanderbilt University, University of Virginia, University of Washington, and Yale University.

\bibliography{all.bib}

\begin{thebibliography}{}
\providecommand\natexlab[1]{#1}
\providecommand\JournalTitle[1]{#1}

\bibitem[{{Abolfathi} {et~al.}(2018){Abolfathi}, {Aguado}, {Aguilar}, {Allende
  Prieto}, {Almeida}, {Ananna}, {Anders}, {Anderson}, {Andrews}, {Anguiano},
  {Arag{\'o}n-Salamanca}, {Argudo-Fern{\'a}ndez}, {Armengaud}, {Ata},
  {Aubourg}, {Avila-Reese}, {Badenes}, {Bailey}, {Balland}, {Barger},
  {Barrera-Ballesteros}, {Bartosz}, {Bastien}, {Bates}, {Baumgarten},
  {Bautista}, {Beaton}, {Beers}, {Belfiore}, {Bender}, {Bernardi}, {Bershady},
  {Beutler}, {Bird}, {Bizyaev}, {Blanc}, {Blanton}, {Blomqvist}, {Bolton},
  {Boquien}, {Borissova}, {Bovy}, {Bradna Diaz}, {Brandt}, {Brinkmann},
  {Brownstein}, {Bundy}, {Burgasser}, {Burtin}, {Busca}, {Ca{\~n}as},
  {Cano-D{\'\i}az}, {Cappellari}, {Carrera}, {Casey}, {Cervantes Sodi}, {Chen},
  {Cherinka}, {Chiappini}, {Choi}, {Chojnowski}, {Chuang}, {Chung}, {Clerc},
  {Cohen}, {Comerford}, {Comparat}, {Correa do Nascimento}, {da Costa},
  {Cousinou}, {Covey}, {Crane}, {Cruz-Gonzalez}, {Cunha}, {da Silva Ilha},
  {Damke}, {Darling}, {Davidson}, {Dawson}, {de Icaza Lizaola}, {de la
  Macorra}, {de la Torre}, {De Lee}, {de Sainte Agathe}, {Deconto Machado},
  {Dell'Agli}, {Delubac}, {Diamond-Stanic}, {Donor}, {Downes}, {Drory}, {du Mas
  des Bourboux}, {Duckworth}, {Dwelly}, {Dyer}, {Ebelke}, {Davis Eigenbrot},
  {Eisenstein}, {Elsworth}, {Emsellem}, {Eracleous}, {Erfanianfar},
  {Escoffier}, {Fan}, {Fern{\'a}ndez Alvar}, {Fernandez-Trincado}, {Fernand o
  Cirolini}, {Feuillet}, {Finoguenov}, {Fleming}, {Font-Ribera}, {Freischlad},
  {Frinchaboy}, {Fu}, {G{\'o}mez Maqueo Chew}, {Galbany}, {Garc{\'\i}a
  P{\'e}rez}, {Garcia-Dias}, {Garc{\'\i}a-Hern{\'a}ndez}, {Garma Oehmichen},
  {Gaulme}, {Gelfand }, {Gil-Mar{\'\i}n}, {Gillespie}, {Goddard}, {Gonz{\'a}lez
  Hern{\'a}ndez}, {Gonzalez-Perez}, {Grabowski}, {Green}, {Grier}, {Gueguen},
  {Guo}, {Guy}, {Hagen}, {Hall}, {Harding}, {Hasselquist}, {Hawley}, {Hayes},
  {Hearty}, {Hekker}, {Hernand ez}, {Hernandez Toledo}, {Hogg},
  {Holley-Bockelmann}, {Holtzman}, {Hou}, {Hsieh}, {Hunt}, {Hutchinson},
  {Hwang}, {Jimenez Angel}, {Johnson}, {Jones}, {J{\"o}nsson}, {Jullo}, {Khan},
  {Kinemuchi}, {Kirkby}, {Kirkpatrick}, {Kitaura}, {Knapp}, {Kneib},
  {Kollmeier}, {Lacerna}, {Lane}, {Lang}, {Law}, {Le Goff}, {Lee}, {Li}, {Li},
  {Lian}, {Liang}, {Lima}, {Lin}, {Long}, {Lucatello}, {Lundgren}, {Mackereth},
  {MacLeod}, {Mahadevan}, {Maia}, {Majewski}, {Manchado}, {Maraston},
  {Mariappan}, {Marques-Chaves}, {Masseron}, {Masters}, {McDermid}, {McGreer},
  {Melendez}, {Meneses-Goytia}, {Merloni}, {Merrifield}, {Meszaros}, {Meza},
  {Minchev}, {Minniti}, {Mueller}, {Muller-Sanchez}, {Muna}, {Mu{\~n}oz},
  {Myers}, {Nair}, {Nand ra}, {Ness}, {Newman}, {Nichol}, {Nidever},
  {Nitschelm}, {Noterdaeme}, {O'Connell}, {Oelkers}, {Oravetz}, {Oravetz},
  {Ort{\'\i}z}, {Osorio}, {Pace}, {Padilla}, {Palanque-Delabrouille},
  {Palicio}, {Pan}, {Pan}, {Parikh}, {P{\^a}ris}, {Park}, {Peirani},
  {Pellejero-Ibanez}, {Penny}, {Percival}, {Perez-Fournon}, {Petitjean},
  {Pieri}, {Pinsonneault}, {Pisani}, {Prada}, {Prakash}, {Queiroz}, {Raddick},
  {Raichoor}, {Barboza Rembold}, {Richstein}, {Riffel}, {Riffel}, {Rix},
  {Robin}, {Rodr{\'\i}guez Torres}, {Rom{\'a}n-Z{\'u}{\~n}iga}, {Ross},
  {Rossi}, {Ruan}, {Ruggeri}, {Ruiz}, {Salvato}, {S{\'a}nchez}, {S{\'a}nchez},
  {Sanchez Almeida}, {S{\'a}nchez-Gallego}, {Santana Rojas}, {Santiago},
  {Schiavon}, {Schimoia}, {Schlafly}, {Schlegel}, {Schneider}, {Schuster},
  {Schwope}, {Seo}, {Serenelli}, {Shen}, {Shen}, {Shetrone}, {Shull}, {Silva
  Aguirre}, {Simon}, {Skrutskie}, {Slosar}, {Smethurst}, {Smith}, {Sobeck},
  {Somers}, {Souter}, {Souto}, {Spindler}, {Stark}, {Stassun}, {Steinmetz},
  {Stello}, {Storchi-Bergmann}, {Streblyanska}, {Stringfellow}, {Su{\'a}rez},
  {Sun}, {Szigeti}, {Taghizadeh-Popp}, {Talbot}, {Tang}, {Tao}, {Tayar},
  {Tembe}, {Teske}, {Thakar}, {Thomas}, {Tissera}, {Tojeiro}, {Tremonti},
  {Troup}, {Urry}, {Valenzuela}, {van den Bosch}, {Vargas-Gonz{\'a}lez},
  {Vargas-Maga{\~n}a}, {Vazquez}, {Villanova}, {Vogt}, {Wake}, {Wang},
  {Weaver}, {Weijmans}, {Weinberg}, {Westfall}, {Whelan}, {Wilcots}, {Wild},
  {Williams}, {Wilson}, {Wood-Vasey}, {Wylezalek}, {Xiao}, {Yan}, {Yang},
  {Ybarra}, {Y{\`e}che}, {Zakamska}, {Zamora}, {Zarrouk}, {Zasowski}, {Zhang},
  {Zhao}, {Zhao}, {Zheng}, {Zheng}, {Zhou}, {Zhu}, {Zinn}, \& {Zou}}]{dr14}
{Abolfathi}, B., {Aguado}, D.~S., {Aguilar}, G., {et~al.} 2018,
  \href{http://dx.doi.org/10.3847/1538-4365/aa9e8a}{\JournalTitle{\apjs}, 235,
  42}

\bibitem[{{Aird} {et~al.}(2018){Aird}, {Coil}, \& {Georgakakis}}]{aird:18}
{Aird}, J., {Coil}, A.~L., \& {Georgakakis}, A. 2018,
  \href{http://dx.doi.org/10.1093/mnras/stx2700}{\JournalTitle{\mnras}, 474,
  1225}

\bibitem[{{Aird} {et~al.}(2019){Aird}, {Coil}, \& {Georgakakis}}]{aird:19}
---. 2019, \href{http://dx.doi.org/10.1093/mnras/stz125}{\JournalTitle{\mnras},
  484, 4360}

\bibitem[{{Aird} {et~al.}(2012){Aird}, {Coil}, {Moustakas}, {Blanton},
  {Burles}, {Cool}, {Eisenstein}, {Smith}, {Wong}, \& {Zhu}}]{aird:12}
{Aird}, J., {Coil}, A.~L., {Moustakas}, J., {et~al.} 2012,
  \href{http://dx.doi.org/10.1088/0004-637X/746/1/90}{\JournalTitle{\apj}, 746,
  90}

\bibitem[{{Baldwin} {et~al.}(1981){Baldwin}, {Phillips}, \&
  {Terlevich}}]{baldwin:81}
{Baldwin}, J.~A., {Phillips}, M.~M., \& {Terlevich}, R. 1981,
  \JournalTitle{\pasp}, 93, 5

\bibitem[{{Bari{\v{s}}i{\'c}} {et~al.}(2017){Bari{\v{s}}i{\'c}}, {van der Wel},
  {Bezanson}, {Pacifici}, {Noeske}, {Mu{\~n}oz-Mateos}, {Franx},
  {Smol{\v{c}}i{\'c}}, {Bell}, {Brammer}, {Calhau}, {Chauk{\'e}}, {van Dokkum},
  {van Houdt}, {Gallazzi}, {Labb{\'e}}, {Maseda}, {Muzzin}, {Sobral},
  {Straatman}, \& {Wu}}]{barisic:17}
{Bari{\v{s}}i{\'c}}, I., {van der Wel}, A., {Bezanson}, R., {et~al.} 2017,
  \href{http://dx.doi.org/10.3847/1538-4357/aa8768}{\JournalTitle{\apj}, 847,
  72}

\bibitem[{{Becker} {et~al.}(1995){Becker}, {White}, \& {Helfand}}]{becker:95}
{Becker}, R.~H., {White}, R.~L., \& {Helfand}, D.~J. 1995,
  \href{http://dx.doi.org/10.1086/176166}{\JournalTitle{\apj}, 450, 559}

\bibitem[{{Bell}(2003)}]{bell:03}
{Bell}, E.~F. 2003,
  \href{http://dx.doi.org/10.1086/367829}{\JournalTitle{\apj}, 586, 794}

\bibitem[{{Best} \& {Heckman}(2012)}]{bestheckman2012}
{Best}, P.~N., \& {Heckman}, T.~M. 2012,
  \href{http://dx.doi.org/10.1111/j.1365-2966.2012.20414.x}{\JournalTitle{\mnras},
  421, 1569}

\bibitem[{{Best} {et~al.}(2005){Best}, {Kauffmann}, {Heckman}, \&
  {Ivezi{\'c}}}]{best:05}
{Best}, P.~N., {Kauffmann}, G., {Heckman}, T.~M., \& {Ivezi{\'c}}, {\v{Z}}.
  2005,
  \href{http://dx.doi.org/10.1111/j.1365-2966.2005.09283.x}{\JournalTitle{\mnras},
  362, 9}

\bibitem[{{Bruzual} \& {Charlot}(2003)}]{bc:03}
{Bruzual}, G., \& {Charlot}, S. 2003, \JournalTitle{\mnras}, 344, 1000

\bibitem[{{Canalizo} \& {Stockton}(2001)}]{canalizo:01}
{Canalizo}, G., \& {Stockton}, A. 2001,
  \href{http://dx.doi.org/10.1086/321520}{\JournalTitle{\apj}, 555, 719}

\bibitem[{{Cappellari} \& {Emsellem}(2004)}]{cappellari:04}
{Cappellari}, M., \& {Emsellem}, E. 2004,
  \href{http://dx.doi.org/10.1086/381875}{\JournalTitle{\pasp}, 116, 138}

\bibitem[{{Chabrier}(2003)}]{chabrier:03}
{Chabrier}, G. 2003, \JournalTitle{\pasp}, 115, 763

\bibitem[{{Chen} {et~al.}(2013){Chen}, {Hickox}, {Alberts}, {Brodwin}, {Jones},
  {Murray}, {Alexander}, {Assef}, {Brown}, {Dey}, {Forman}, {Gorjian},
  {Goulding}, {Le Floc'h}, {Jannuzi}, {Mullaney}, \& {Pope}}]{chen:13}
{Chen}, C.-T.~J., {Hickox}, R.~C., {Alberts}, S., {et~al.} 2013,
  \href{http://dx.doi.org/10.1088/0004-637X/773/1/3}{\JournalTitle{\apj}, 773,
  3}

\bibitem[{{Croton} {et~al.}(2006){Croton}, {Springel}, {White}, {De Lucia},
  {Frenk}, {Gao}, {Jenkins}, {Kauffmann}, {Navarro}, \& {Yoshida}}]{croton:06}
{Croton}, D.~J., {Springel}, V., {White}, S.~D.~M., {et~al.} 2006,
  \href{http://dx.doi.org/10.1111/j.1365-2966.2005.09675.x}{\JournalTitle{\mnras},
  365, 11}

\bibitem[{{Davies} {et~al.}(2007){Davies}, {M{\"u}ller S{\'a}nchez}, {Genzel},
  {Tacconi}, {Hicks}, {Friedrich}, \& {Sternberg}}]{davies:07}
{Davies}, R.~I., {M{\"u}ller S{\'a}nchez}, F., {Genzel}, R., {et~al.} 2007,
  \href{http://dx.doi.org/10.1086/523032}{\JournalTitle{\apj}, 671, 1388}

\bibitem[{{Dawson} {et~al.}(2013){Dawson}, {Schlegel}, {Ahn}, {Anderson},
  {Aubourg}, {Bailey}, {Barkhouser}, {Bautista}, {Beifiori}, {Berlind},
  {Bhardwaj}, {Bizyaev}, {Blake}, {Blanton}, {Blomqvist}, {Bolton}, {Borde},
  {Bovy}, {Brandt}, {Brewington}, {Brinkmann}, {Brown}, {Brownstein}, {Bundy},
  {Busca}, {Carithers}, {Carnero}, {Carr}, {Chen}, {Comparat}, {Connolly},
  {Cope}, {Croft}, {Cuesta}, {da Costa}, {Davenport}, {Delubac}, {de Putter},
  {Dhital}, {Ealet}, {Ebelke}, {Eisenstein}, {Escoffier}, {Fan}, {Filiz Ak},
  {Finley}, {Font-Ribera}, {G{\'e}nova-Santos}, {Gunn}, {Guo}, {Haggard},
  {Hall}, {Hamilton}, {Harris}, {Harris}, {Ho}, {Hogg}, {Holder}, {Honscheid},
  {Huehnerhoff}, {Jordan}, {Jordan}, {Kauffmann}, {Kazin}, {Kirkby}, {Klaene},
  {Kneib}, {Le Goff}, {Lee}, {Long}, {Loomis}, {Lundgren}, {Lupton}, {Maia},
  {Makler}, {Malanushenko}, {Malanushenko}, {Mandelbaum}, {Manera}, {Maraston},
  {Margala}, {Masters}, {McBride}, {McDonald}, {McGreer}, {McMahon}, {Mena},
  {Miralda-Escud{\'e}}, {Montero-Dorta}, {Montesano}, {Muna}, {Myers},
  {Naugle}, {Nichol}, {Noterdaeme}, {Nuza}, {Olmstead}, {Oravetz}, {Oravetz},
  {Owen}, {Padmanabhan}, {Palanque-Delabrouille}, {Pan}, {Parejko},
  {P{\^a}ris}, {Percival}, {P{\'e}rez-Fournon}, {P{\'e}rez-R{\`a}fols},
  {Petitjean}, {Pfaffenberger}, {Pforr}, {Pieri}, {Prada}, {Price-Whelan},
  {Raddick}, {Rebolo}, {Rich}, {Richards}, {Rockosi}, {Roe}, {Ross}, {Ross},
  {Rossi}, {Rubi{\~n}o-Martin}, {Samushia}, {S{\'a}nchez}, {Sayres}, {Schmidt},
  {Schneider}, {Sc{\'o}ccola}, {Seo}, {Shelden}, {Sheldon}, {Shen}, {Shu},
  {Slosar}, {Smee}, {Snedden}, {Stauffer}, {Steele}, {Strauss}, {Streblyanska},
  {Suzuki}, {Swanson}, {Tal}, {Tanaka}, {Thomas}, {Tinker}, {Tojeiro},
  {Tremonti}, {Vargas Maga{\~n}a}, {Verde}, {Viel}, {Wake}, {Watson}, {Weaver},
  {Weinberg}, {Weiner}, {West}, {White}, {Wood-Vasey}, {Yeche}, {Zehavi},
  {Zhao}, \& {Zheng}}]{dawson:13}
{Dawson}, K.~S., {Schlegel}, D.~J., {Ahn}, C.~P., {et~al.} 2013,
  \href{http://dx.doi.org/10.1088/0004-6256/145/1/10}{\JournalTitle{\aj}, 145,
  10}

\bibitem[{{de Vries} {et~al.}(2007){de Vries}, {Hodge}, {Becker}, {White}, \&
  {Helfand}}]{deVries:07}
{de Vries}, W.~H., {Hodge}, J.~A., {Becker}, R.~H., {White}, R.~L., \&
  {Helfand}, D.~J. 2007,
  \href{http://dx.doi.org/10.1086/518866}{\JournalTitle{\aj}, 134, 457}

\bibitem[{{Fabian}(2012)}]{fabian:12}
{Fabian}, A.~C. 2012,
  \href{http://dx.doi.org/10.1146/annurev-astro-081811-125521}{\JournalTitle{\araa},
  50, 455}

\bibitem[{{Goulding} \& {Alexander}(2009)}]{goulding:09}
{Goulding}, A.~D., \& {Alexander}, D.~M. 2009,
  \href{http://dx.doi.org/10.1111/j.1365-2966.2009.15194.x}{\JournalTitle{\mnras},
  398, 1165}

\bibitem[{{Greene} {et~al.}(2012){Greene}, {Murphy}, {Comerford}, {Gebhardt},
  \& {Adams}}]{greene:12}
{Greene}, J.~E., {Murphy}, J.~D., {Comerford}, J.~M., {Gebhardt}, K., \&
  {Adams}, J.~J. 2012,
  \href{http://dx.doi.org/10.1088/0004-637X/750/1/32}{\JournalTitle{\apj}, 750,
  32}

\bibitem[{{H{\"a}ring} \& {Rix}(2004)}]{haeringrix:04}
{H{\"a}ring}, N., \& {Rix}, H.-W. 2004,
  \href{http://dx.doi.org/10.1086/383567}{\JournalTitle{\apjl}, 604, L89}

\bibitem[{{Heckman} \& {Best}(2014)}]{heckmanbest:14}
{Heckman}, T.~M., \& {Best}, P.~N. 2014,
  \href{http://dx.doi.org/10.1146/annurev-astro-081913-035722}{\JournalTitle{\araa},
  52, 589}

\bibitem[{{Hine} \& {Longair}(1979)}]{hinelongair1979}
{Hine}, R.~G., \& {Longair}, M.~S. 1979,
  \href{http://dx.doi.org/10.1093/mnras/188.1.111}{\JournalTitle{\mnras}, 188,
  111}

\bibitem[{{Ho}(2008)}]{ho:08}
{Ho}, L.~C. 2008,
  \href{http://dx.doi.org/10.1146/annurev.astro.45.051806.110546}{\JournalTitle{\araa},
  46, 475}

\bibitem[{{Hunt} {et~al.}(2018){Hunt}, {Bezanson}, {Greene}, {Spilker},
  {Suess}, {Kriek}, {Narayanan}, {Feldmann}, {van der Wel}, \&
  {Pattarakijwanich}}]{hunt:18}
{Hunt}, Q., {Bezanson}, R., {Greene}, J.~E., {et~al.} 2018,
  \href{http://dx.doi.org/10.3847/2041-8213/aaca9a}{\JournalTitle{\apjl}, 860,
  L18}

\bibitem[{{Jones} {et~al.}(2016){Jones}, {Hickox}, {Black}, {Hainline},
  {DiPompeo}, \& {Goulding}}]{jones:16}
{Jones}, M.~L., {Hickox}, R.~C., {Black}, C.~S., {et~al.} 2016,
  \href{http://dx.doi.org/10.3847/0004-637X/826/1/12}{\JournalTitle{\apj}, 826,
  12}

\bibitem[{{Juneau} {et~al.}(2011){Juneau}, {Dickinson}, {Alexander}, \&
  {Salim}}]{juneau:11}
{Juneau}, S., {Dickinson}, M., {Alexander}, D.~M., \& {Salim}, S. 2011,
  \href{http://dx.doi.org/10.1088/0004-637X/736/2/104}{\JournalTitle{\apj},
  736, 104}

\bibitem[{{Kauffmann} {et~al.}(2003){Kauffmann}, {Heckman}, {Tremonti},
  {Brinchmann}, {Charlot}, \& {ETAL}}]{kauffmann:03}
{Kauffmann}, G., {Heckman}, T.~M., {Tremonti}, C., {et~al.} 2003,
  \JournalTitle{\mnras}, 346, 1055

\bibitem[{{Kewley} {et~al.}(2013){Kewley}, {Maier}, {Yabe}, {Ohta}, {Akiyama},
  {Dopita}, \& {Yuan}}]{kewley:13}
{Kewley}, L.~J., {Maier}, C., {Yabe}, K., {et~al.} 2013,
  \href{http://dx.doi.org/10.1088/2041-8205/774/1/L10}{\JournalTitle{\apjl},
  774, L10}

\bibitem[{{Kriek} {et~al.}(2009){Kriek}, {van Dokkum}, {Labb{\'e}}, {Franx},
  {Illingworth}, {Marchesini}, \& {Quadri}}]{kriek:09}
{Kriek}, M., {van Dokkum}, P.~G., {Labb{\'e}}, I., {et~al.} 2009,
  \href{http://dx.doi.org/10.1088/0004-637X/700/1/221}{\JournalTitle{\apj},
  700, 221}

\bibitem[{{Kriek} {et~al.}(2010){Kriek}, {Labb{\'e}}, {Conroy}, {Whitaker},
  {van Dokkum}, {Brammer}, {Franx}, {Illingworth}, {Marchesini}, {Muzzin},
  {Quadri}, \& {Rudnick}}]{kriek:10}
{Kriek}, M., {Labb{\'e}}, I., {Conroy}, C., {et~al.} 2010,
  \href{http://dx.doi.org/10.1088/2041-8205/722/1/L64}{\JournalTitle{\apjl},
  722, L64}

\bibitem[{{Leauthaud} {et~al.}(2016){Leauthaud}, {Bundy}, {Saito}, {Tinker},
  {Maraston}, {Tojeiro}, {Huang}, {Brownstein}, {Schneider}, \&
  {Thomas}}]{leauthaud:16}
{Leauthaud}, A., {Bundy}, K., {Saito}, S., {et~al.} 2016,
  \href{http://dx.doi.org/10.1093/mnras/stw117}{\JournalTitle{\mnras}, 457,
  4021}

\bibitem[{{Liu} {et~al.}(2013){Liu}, {Zakamska}, {Greene}, {Nesvadba}, \&
  {Liu}}]{liu:13}
{Liu}, G., {Zakamska}, N.~L., {Greene}, J.~E., {Nesvadba}, N. P.~H., \& {Liu},
  X. 2013,
  \href{http://dx.doi.org/10.1093/mnras/stt1755}{\JournalTitle{\mnras}, 436,
  2576}

\bibitem[{{Liu} {et~al.}(2009){Liu}, {Zakamska}, {Greene}, {Strauss}, {Krolik},
  \& {Heckman}}]{liu:09}
{Liu}, X., {Zakamska}, N.~L., {Greene}, J.~E., {et~al.} 2009,
  \href{http://dx.doi.org/10.1088/0004-637X/702/2/1098}{\JournalTitle{\apj},
  702, 1098}

\bibitem[{{Mandelbaum} {et~al.}(2009){Mandelbaum}, {Li}, {Kauffmann}, \&
  {White}}]{mandelbaumetal2009}
{Mandelbaum}, R., {Li}, C., {Kauffmann}, G., \& {White}, S. D.~M. 2009,
  \href{http://dx.doi.org/10.1111/j.1365-2966.2008.14235.x}{\JournalTitle{\mnras},
  393, 377}

\bibitem[{{Moustakas} {et~al.}(2006){Moustakas}, {Kennicutt}, \&
  {Tremonti}}]{moustakas:06}
{Moustakas}, J., {Kennicutt}, Robert~C., J., \& {Tremonti}, C.~A. 2006,
  \href{http://dx.doi.org/10.1086/500964}{\JournalTitle{\apj}, 642, 775}

\bibitem[{{Muzzin} {et~al.}(2013){Muzzin}, {Marchesini}, {Stefanon}, {Franx},
  {Milvang-Jensen}, {Dunlop}, {Fynbo}, {Brammer}, {Labb{\'e}}, \& {van
  Dokkum}}]{muzzin:13ultravista}
{Muzzin}, A., {Marchesini}, D., {Stefanon}, M., {et~al.} 2013,
  \href{http://dx.doi.org/10.1088/0067-0049/206/1/8}{\JournalTitle{\apjs}, 206,
  8}

\bibitem[{{Pattarakijwanich} {et~al.}(2014){Pattarakijwanich}, {Strauss}, {Ho},
  \& {Ross}}]{pattarakijwanich:14}
{Pattarakijwanich}, P., {Strauss}, M.~A., {Ho}, S., \& {Ross}, N.~P. 2014,
  \JournalTitle{arXiv:1410.7394},
  \href{http://arxiv.org/abs/1410.7394}{{\sffamily arXiv:1410.7394}}

\bibitem[{{Reyes} {et~al.}(2008){Reyes}, {Zakamska}, {Strauss}, {Green},
  {Krolik}, {Shen}, {Richards}, {Anderson}, \& {Schneider}}]{reyes:08}
{Reyes}, R., {Zakamska}, N.~L., {Strauss}, M.~A., {et~al.} 2008,
  \href{http://dx.doi.org/10.1088/0004-6256/136/6/2373}{\JournalTitle{\aj},
  136, 2373}

\bibitem[{{Sabater} {et~al.}(2013){Sabater}, {Best}, \&
  {Argudo-Fern{\'a}ndez}}]{sabater:13}
{Sabater}, J., {Best}, P.~N., \& {Argudo-Fern{\'a}ndez}, M. 2013,
  \href{http://dx.doi.org/10.1093/mnras/sts675}{\JournalTitle{\mnras}, 430,
  638}

\bibitem[{{S{\'a}nchez-Bl{\'a}zquez} {et~al.}(2006){S{\'a}nchez-Bl{\'a}zquez},
  {Peletier}, {Jim{\'e}nez-Vicente}, {Cardiel}, {Cenarro},
  {Falc{\'o}n-Barroso}, {Gorgas}, {Selam}, \& {Vazdekis}}]{miles}
{S{\'a}nchez-Bl{\'a}zquez}, P., {Peletier}, R.~F., {Jim{\'e}nez-Vicente}, J.,
  {et~al.} 2006,
  \href{http://dx.doi.org/10.1111/j.1365-2966.2006.10699.x}{\JournalTitle{\mnras},
  371, 703}

\bibitem[{{Sanders} {et~al.}(1988){Sanders}, {Soifer}, {Elias}, {Madore},
  {Matthews}, {Neugebauer}, \& {Scoville}}]{sanders:88}
{Sanders}, D.~B., {Soifer}, B.~T., {Elias}, J.~H., {et~al.} 1988,
  \JournalTitle{\apj}, 325, 74

\bibitem[{{Shakura} \& {Sunyaev}(1973)}]{shakurasunyaev1973}
{Shakura}, N.~I., \& {Sunyaev}, R.~A. 1973, \JournalTitle{\aap}, 500, 33

\bibitem[{{Shapley} {et~al.}(2005){Shapley}, {Coil}, {Ma}, \&
  {Bundy}}]{shapley:05}
{Shapley}, A.~E., {Coil}, A.~L., {Ma}, C.-P., \& {Bundy}, K. 2005,
  \href{http://dx.doi.org/10.1086/497630}{\JournalTitle{\apj}, 635, 1006}

\bibitem[{{Shapley} {et~al.}(2015){Shapley}, {Reddy}, {Kriek}, {Freeman},
  {Sanders}, {Siana}, {Coil}, {Mobasher}, {Shivaei}, {Price}, \& {de
  Groot}}]{shapley:15}
{Shapley}, A.~E., {Reddy}, N.~A., {Kriek}, M., {et~al.} 2015,
  \href{http://dx.doi.org/10.1088/0004-637X/801/2/88}{\JournalTitle{\apj}, 801,
  88}

\bibitem[{{Silk} \& {Rees}(1998)}]{silk:98}
{Silk}, J., \& {Rees}, M.~J. 1998, \JournalTitle{\aap}, 331, L1

\bibitem[{{Silverman} {et~al.}(2009){Silverman}, {Lamareille}, {Maier},
  {Lilly}, {Mainieri}, {Brusa}, {Cappelluti}, {Hasinger}, {Zamorani},
  {Scodeggio}, {Bolzonella}, {Contini}, {Carollo}, {Jahnke}, {Kneib}, {Le
  F{\`e}vre}, {Merloni}, {Bardelli}, {Bongiorno}, {Brunner}, {Caputi},
  {Civano}, {Comastri}, {Coppa}, {Cucciati}, {de la Torre}, {de Ravel},
  {Elvis}, {Finoguenov}, {Fiore}, {Franzetti}, {Garilli}, {Gilli}, {Iovino},
  {Kampczyk}, {Knobel}, {Kova{\v{c}}}, {Le Borgne}, {Le Brun}, {Mignoli},
  {Pello}, {Peng}, {Perez Montero}, {Ricciardelli}, {Tanaka}, {Tasca},
  {Tresse}, {Vergani}, {Vignali}, {Zucca}, {Bottini}, {Cappi}, {Cassata},
  {Fumana}, {Griffiths}, {Kartaltepe}, {Koekemoer}, {Marinoni}, {McCracken},
  {Memeo}, {Meneux}, {Oesch}, {Porciani}, \& {Salvato}}]{silverman:09}
{Silverman}, J.~D., {Lamareille}, F., {Maier}, C., {et~al.} 2009,
  \href{http://dx.doi.org/10.1088/0004-637X/696/1/396}{\JournalTitle{\apj},
  696, 396}

\bibitem[{{Smol{\v{c}}i{\'c}} {et~al.}(2017){Smol{\v{c}}i{\'c}}, {Novak},
  {Bondi}, {Ciliegi}, {Mooley}, {Schinnerer}, {Zamorani}, {Navarrete},
  {Bourke}, {Karim}, {Vardoulaki}, {Leslie}, {Delhaize}, {Carilli}, {Myers},
  {Baran}, {Delvecchio}, {Miettinen}, {Banfield}, {Balokovi{\'c}}, {Bertoldi},
  {Capak}, {Frail}, {Hallinan}, {Hao}, {Herrera Ruiz}, {Horesh}, {Ilbert},
  {Intema}, {Jeli{\'c}}, {Kl{\"o}ckner}, {Krpan}, {Kulkarni}, {McCracken},
  {Laigle}, {Middleberg}, {Murphy}, {Sargent}, {Scoville}, \&
  {Sheth}}]{smolcic:17}
{Smol{\v{c}}i{\'c}}, V., {Novak}, M., {Bondi}, M., {et~al.} 2017,
  \href{http://dx.doi.org/10.1051/0004-6361/201628704}{\JournalTitle{\aap},
  602, A1}

\bibitem[{{Springel} {et~al.}(2005){Springel}, {White}, {Jenkins}, {Frenk},
  {Yoshida}, {Gao}, {Navarro}, {Thacker}, {Croton}, {Helly}, {Peacock}, {Cole},
  {Thomas}, {Couchman}, {Evrard}, {Colberg}, \& {Pearce}}]{springel:05}
{Springel}, V., {White}, S.~D.~M., {Jenkins}, A., {et~al.} 2005,
  \href{http://dx.doi.org/10.1038/nature03597}{\JournalTitle{\nat}, 435, 629}

\bibitem[{{Straatman} {et~al.}(2018){Straatman}, {van der Wel}, {Bezanson},
  {Pacifici}, {Gallazzi}, {Wu}, {Noeske}, {Bari{\v s}i{\'c}}, {Bell},
  {Brammer}, {Calhau}, {Chauke}, {Franx}, {van Houdt}, {Labb{\'e}}, {Maseda},
  {Mu{\~n}oz-Mateos}, {Muzzin}, {van de Sande}, {Sobral}, \&
  {Spilker}}]{straatman:18}
{Straatman}, C.~M.~S., {van der Wel}, A., {Bezanson}, R., {et~al.} 2018,
  \href{http://dx.doi.org/10.3847/1538-4365/aae37a}{\JournalTitle{\apjs}, 239,
  27}

\bibitem[{{Suess} {et~al.}(2017){Suess}, {Bezanson}, {Spilker}, {Kriek},
  {Greene}, {Feldmann}, {Hunt}, \& {Narayanan}}]{suess:17}
{Suess}, K.~A., {Bezanson}, R., {Spilker}, J.~S., {et~al.} 2017,
  \href{http://dx.doi.org/10.3847/2041-8213/aa85dc}{\JournalTitle{\apjl}, 846,
  L14}

\bibitem[{{Thomas} {et~al.}(2005){Thomas}, {Maraston}, {Bender}, \& {Mendes de
  Oliveira}}]{thomas:05}
{Thomas}, D., {Maraston}, C., {Bender}, R., \& {Mendes de Oliveira}, C. 2005,
  \href{http://dx.doi.org/10.1086/426932}{\JournalTitle{\apj}, 621, 673}

\bibitem[{{Tremonti} {et~al.}(2007){Tremonti}, {Moustakas}, \&
  {Diamond-Stanic}}]{tremonti:07}
{Tremonti}, C.~A., {Moustakas}, J., \& {Diamond-Stanic}, A.~M. 2007,
  \href{http://dx.doi.org/10.1086/520083}{\JournalTitle{\apjl}, 663, L77}

\bibitem[{{Trump} {et~al.}(2015){Trump}, {Sun}, {Zeimann}, {Luck}, {Bridge},
  {Grier}, {Hagen}, {Juneau}, {Montero-Dorta}, {Rosario}, {Brandt},
  {Ciardullo}, \& {Schneider}}]{trump:15}
{Trump}, J.~R., {Sun}, M., {Zeimann}, G.~R., {et~al.} 2015,
  \href{http://dx.doi.org/10.1088/0004-637X/811/1/26}{\JournalTitle{\apj}, 811,
  26}

\bibitem[{{van der Wel} {et~al.}(2016){van der Wel}, {Noeske}, {Bezanson},
  {Pacifici}, {Gallazzi}, {Franx}, {Mu{\~n}oz-Mateos}, {Bell}, {Brammer},
  {Charlot}, {Chauk{\'e}}, {Labb{\'e}}, {Maseda}, {Muzzin}, {Rix}, {Sobral},
  {van de Sande}, {van Dokkum}, {Wild}, \& {Wolf}}]{wel:16}
{van der Wel}, A., {Noeske}, K., {Bezanson}, R., {et~al.} 2016,
  \href{http://dx.doi.org/10.3847/0067-0049/223/2/29}{\JournalTitle{\apjs},
  223, 29}

\bibitem[{{Vazdekis} {et~al.}(2010){Vazdekis}, {S{\'a}nchez-Bl{\'a}zquez},
  {Falc{\'o}n-Barroso}, {Cenarro}, {Beasley}, {Cardiel}, {Gorgas}, \&
  {Peletier}}]{vazdekis:2010}
{Vazdekis}, A., {S{\'a}nchez-Bl{\'a}zquez}, P., {Falc{\'o}n-Barroso}, J.,
  {et~al.} 2010,
  \href{http://dx.doi.org/10.1111/j.1365-2966.2010.16407.x}{\JournalTitle{\mnras},
  404, 1639}

\bibitem[{{Wild} {et~al.}(2016){Wild}, {Almaini}, {Dunlop}, {Simpson},
  {Rowlands}, {Bowler}, {Maltby}, \& {McLure}}]{wild:16}
{Wild}, V., {Almaini}, O., {Dunlop}, J., {et~al.} 2016,
  \href{http://dx.doi.org/10.1093/mnras/stw1996}{\JournalTitle{\mnras}, 463,
  832}

\bibitem[{{Wild} {et~al.}(2010){Wild}, {Heckman}, \& {Charlot}}]{wild:10}
{Wild}, V., {Heckman}, T., \& {Charlot}, S. 2010,
  \href{http://dx.doi.org/10.1111/j.1365-2966.2010.16536.x}{\JournalTitle{\mnras},
  405, 933}

\bibitem[{{Yang} {et~al.}(2017){Yang}, {Chen}, {Vito}, {Brandt}, {Alexander},
  {Luo}, {Sun}, {Xue}, {Bauer}, {Koekemoer}, {Lehmer}, {Liu}, {Schneider},
  {Shemmer}, {Trump}, {Vignali}, \& {Wang}}]{yang:17}
{Yang}, G., {Chen}, C. T.~J., {Vito}, F., {et~al.} 2017,
  \href{http://dx.doi.org/10.3847/1538-4357/aa7564}{\JournalTitle{\apj}, 842,
  72}

\bibitem[{{Yang} {et~al.}(2018){Yang}, {Brandt}, {Vito}, {Chen}, {Trump},
  {Luo}, {Sun}, {Xue}, {Koekemoer}, {Schneider}, {Vignali}, \&
  {Wang}}]{yang:18}
{Yang}, G., {Brandt}, W.~N., {Vito}, F., {et~al.} 2018,
  \href{http://dx.doi.org/10.1093/mnras/stx2805}{\JournalTitle{\mnras}, 475,
  1887}

\bibitem[{{Zakamska} {et~al.}(2006){Zakamska}, {Strauss}, {Krolik}, {Ridgway},
  {Schmidt}, {Smith}, {Heckman}, {Schneider}, {Hao}, \&
  {Brinkmann}}]{zakamska:06}
{Zakamska}, N.~L., {Strauss}, M.~A., {Krolik}, J.~H., {et~al.} 2006,
  \href{http://dx.doi.org/10.1086/506986}{\JournalTitle{\aj}, 132, 1496}

\end{thebibliography}

\end{document}